\documentclass[11pt]
{article} 
\usepackage{graphicx}
\usepackage{amsmath}
\usepackage{bbm}
\usepackage{cancel}
\usepackage[normalem]{ulem}

\usepackage{amsfonts}
\usepackage{amssymb}
\usepackage{amsmath}
\usepackage{latexsym}
\usepackage{xcolor}
\usepackage{url}
\usepackage{ntheorem}
\usepackage{braket}

\DeclareFontFamily{OT1}{rsfs}{}
\DeclareFontShape{OT1}{rsfs}{m}{n}{ <-7> rsfs5 <7-10> rsfs7 <10->
rsfs10}{} \DeclareMathAlphabet{\mycal}{OT1}{rsfs}{m}{n}

\newcommand{\divrm}{{\mathrm{div}}}
\newcommand{\curl}{{\mathrm{curl}}}

\newcounter{mnotecount}[section]

\renewcommand{\themnotecount}{\thesection.\arabic{mnotecount}}
\newcommand{\mnote}[1]
{\protect{\stepcounter{mnotecount}}$^{\mbox{\footnotesize
$
\bullet$\themnotecount}}$ \marginpar{
\raggedright\tiny\em
$\!\!\!\!\!\!\,\bullet$\themnotecount: #1} }

\newtheorem{theorem}{\sc  Theorem\rm}[section]

\newtheorem{lemma}[theorem]{\sc Lemma\rm}

\newcommand{\eeal}[1]{\label{#1}\end{eqnarray}}
\newcommand{\bed}{\begin{deqarr}}
\newcommand{\eed}{\end{deqarr}}
\newcommand{\bedl}[1]{\begin{deqarr}\label{#1}}
\newcommand{\eedl}[2]{\arrlabel{#1}\label{#2}\end{deqarr}}

\newcommand{\mcU}{{\mycal U}}

\newcommand{\bel}[1]{\begin{equation}\label{#1}}
\newcommand{\bea}{\begin{eqnarray}}
\newcommand{\bean}{\begin{eqnarray}\nonumber}
\newcommand{\beal}[1]{\begin{eqnarray}\label{#1}}
\newcommand{\eea}{\end{eqnarray}}

\newcommand{\nn}{\nonumber}

\newcommand{\Eqs}[2]{Equations~\eq{#1}-\eq{#2}}

\newcommand{\qed}{\hfill $\Box$ \medskip}
\newcommand{\proof}{\noindent {\sc Proof:\ }}
\newcommand{\be}{\begin{equation}}
\newcommand{\eeq}{\end{equation}}
\newcommand{\ee}{\end{equation}}
\newcommand{\beqa}{\begin{eqnarray}}
\newcommand{\eeqa}{\end{eqnarray}}
\newcommand{\beqan}{\begin{eqnarray*}}
\newcommand{\eeqan}{\end{eqnarray*}}
\newcommand{\ba}{\begin{array}}
\newcommand{\ea}{\end{array}}

\newcommand{\const}{\mbox{\rm const}} 

\newcommand{\R}{\mathbb R}

\newcommand{\eq}[1]{(\ref{#1})}

\newcommand{\beqar}{\begin{deqarr}}
\newcommand{\eeqar}{\end{deqarr}}

\newcommand{\beaa}{\begin{eqnarray*}}
\newcommand{\eeaa}{\end{eqnarray*}}

{\catcode `\@=11 \global\let\AddToReset=\@addtoreset}
\AddToReset{equation}{section}

\begin{document}
\title{Shielding linearised-gravity\protect\thanks{Preprint UWThPh-2016-31}}
\author{Robert Beig\thanks{
{\sc Email} \protect\url{robert.beig@univie.ac.at}},
Piotr T. Chru\'{s}ciel\thanks{
{\sc Email} \protect\url{piotr.chrusciel@univie.ac.at}, {\sc URL} \protect\url{homepage.univie.ac.at/piotr.chrusciel}}
\\
University of Vienna
}
\maketitle

\begin{abstract}
We present an elementary argument that one can shield linearised gravitational fields using linearised gravitational fields. This is done by using third-order potentials for the metric, which avoids the need to solve singular equations in shielding or gluing constructions for the linearised metric.
\end{abstract}

\tableofcontents

\section{Introduction}
 \label{s23XI16.1}

A fundamental property of Newtonian gravity is that the gravitational field cannot be localised in a bounded region. This is a simple consequence of the equation
$$
 \Delta \phi = 4 \pi G \rho
 \,,
$$
where $\phi$ is the gravitational potential, $G$ is Newton's constant and $\rho$ is the matter density:  The requirement that $\rho \ge 0$, and the asymptotic behaviour $-M/r$ of $\phi$, where $M$ is the total mass, implies that $\phi$ vanishes at large distances along a curve extending to infinity if and only if there is no matter whatsoever and $\phi \equiv 0$.
It is therefore extremely surprising that \emph{in general relativity, gravitational fields can be shielded away by gravitational fields}, as proved recently in a remarkable paper by  Carlotto and Schoen~\cite{CarlottoSchoen}.

Since Newtonian gravity is part of the weak-field limit of general relativity (indeed, this is weak-field GR with small velocities),
one wonders if a similar screening can occur for linearised relativity. As it turns out, the analysis of Carlotto and Schoen can be readily generalised to linearised gravitational fields on cone-like sets as considered in~\cite{CarlottoSchoen} (compare~\cite{ChDelayExoticAERiemannian}). This,  however, requires sophistical mathematical machinery which imposes restrictions to the sets considered
and, as an intermediate step, uses solutions blowing-up at the relevant boundaries, which leads to difficulties when trying to implement the method numerically. The object of this note is to point out an alternative elementary method to
perform gluings, or achieve screening of linearised gravity by linearised gravitational fields near a Minkowski background.
In particular, we give here a very simple proof that \emph{at any given time $t$, and given any open set $\Omega\subset \R^3$, every linearised vacuum gravitational field $h_{\mu\nu}$ on $\{t\}\times \R^3$ can be deformed to a new linearised vacuum field $\tilde h_{\mu\nu}$ so that $\tilde h_{\mu\nu}$ coincides with $ h_{\mu\nu}$ on $\Omega$ and vanishes outside a slightly larger set}. In other words, the gravitational field has been screened away outside of $\Omega$, and this by using gravitational fields only: no matter fields, whether with positive or negative density, are needed.

We emphasise that the construction of Carlotto-Schoen switches-off the gravitational field in sets which have a cone-like structure, whether in the linearised case or in the full treatment. In our approach no restrictions on the geometry of $\Omega$ occur, so that the screening can be done near any set.

Our construction is likely to be useful for the numerical construction of initial data sets with interesting properties, by providing an efficient way of making gluings in the far-away zone, where nonlinear corrections become inessential. Here, as already pointed out, both the Corvino-Schoen and the Carlotto-Schoen gluings require solving elliptic equations in spaces of functions which are singular at the boundary of the gluing region (see~\cite{CorvinoPollack,ChBourbaki} for a review), while our gluings are performed by explicit elementary integrations (\eq{17XII16.14} below), multiplication by a cut-off function, and applying derivatives, once the metric has been put into transverse and traceless (TT)-gauge.

The above leads one naturally to ask similar questions for electric and magnetic fields. Here we provide a simple proof that Maxwell fields can be shielded by Maxwell fields.  Last but not least, we show how to perform the screening in practice, in that we prove that all solutions of sourceless Maxwell equations in a bounded space-time region can be realised by manipulating charges and currents in an enclosing bounded region.

\section{Shielding linearised gravity}
 \label{ss12XII16.2}
Consider $\R^{{3}+1}$ with a metric which,
in the natural coordinates on $\R^{3+1}$, takes the form
\bel{+ExSp1} g_{\mu\nu}=\eta_{\mu\nu}+h_{\mu\nu}\,,
\ee
where $\eta$ denotes the Minkowski metric. Suppose
that there exists a small constant $\epsilon$ such that we have
\bel{+ExSp2} |h_{\mu\nu}|\,,\ |\partial_\sigma h_{\mu\nu}| \,,\
|\partial_\sigma\partial_\rho h_{\mu\nu}| = O(\epsilon)
 \,.
\ee
If we use the metric $\eta$ to raise and
lower indices one has
\bea
R_{\beta\delta}   &=&\frac 12
\left[\partial_\alpha\{\partial_\beta h^\alpha{}_{\delta} +
\partial_\delta h^\alpha{}_{\beta}- \partial^\alpha
h_{\beta\delta}\} - \partial_\delta\partial_\beta
h^\alpha{}_{\alpha} \right] +O(\epsilon^2)
\,.
\eeal{+ExSp3}
Coordinate transformations $x^\mu\mapsto x^\mu + \zeta^\mu$, with
\bel{12XII16.1}
  |\zeta_{\mu}| \,,
 \quad |\partial_\sigma \zeta_{\mu}|
  \,, \quad
 |\partial_\sigma\partial_\rho \zeta_{\mu}|
  \,, \quad
 |\partial_\sigma\partial_\rho \partial_\nu \zeta_{\mu}| = O( \epsilon)
\,,
\ee
preserve \eq{+ExSp2}, and lead to the gauge-freedom
\bel{12XII16.2}
 h_{\mu\nu} \mapsto h_{\mu\nu} + \partial_\mu \zeta_\nu + \partial_\nu \zeta_\mu
  \,.
\ee
Imposing the wave-coordinates condition up to $O(\epsilon^2)$ terms,
\bea
  \Box_g x^\alpha = O(\epsilon^2)
 \,,
\label{+ExSp21}
\eea
leads to
\bel{+ExSp17}\partial_\beta
h^{\beta}{}_\alpha=\frac 12 \partial_\alpha h^\beta{}_\beta+
O(\epsilon^2)
 \,,
\ee
as well as
\bea
 R_{\beta\delta}
 &=&-\frac 12 \Box_\eta h_{\beta\delta}
 +O(\epsilon^2)
  \,.
\eeal{+ExSp23}

\subsection{The Cauchy problem for linearised gravity}
 \label{ss12XII16.1}

In what follows we ignore all $O(\epsilon^2)$-terms in the equations above and consider the theory of a tensor field $h_{\mu\nu}$ with the gauge-freedom \eq{12XII16.2} and satisfying the equations
\bea
0 =
\partial_\alpha\{\partial_\beta h^\alpha{}_{\delta} +
\partial_\delta h^\alpha{}_{\beta}- \partial^\alpha
h_{\beta\delta}\} - \partial_\delta\partial_\beta
h^\alpha{}_{\alpha}
\,.
\eeal{++ExSp3}
Solving the following wave equation
$$
 \Box \zeta_\alpha = - \partial_\beta h^{\beta}{}_\alpha+\frac 12 \partial_\alpha h^\beta{}_\beta
 \,,
$$
where $\Box \equiv \Box_\eta $ is the wave-operator of the Minkowski metric, and performing \eq{12XII16.2}
leads to a new tensor $h_{\mu\nu}$, still denoted by the same symbol, such that
\bel{++ExSp17}\partial_\beta
 h^{\beta}{}_\alpha=\frac 12 \partial_\alpha h^\beta{}_\beta
 \,,
\ee
together with the usual wave equation for $h$:
\bea
 \Box  h_{\beta\delta}
  =0
  \,.
\eeal{++ExSp23}
Solutions of this last equation are in one-to-one correspondence with their Cauchy data at $t=0$. However, those data are not arbitrary, which can be seen as follows: \Eqs{++ExSp17}{++ExSp23} imply
\bel{12XII16.4}
 \Box ( \partial_\beta h^{\beta}{}_\alpha- \frac 12 \partial_\alpha h^\beta{}_\beta) = 0
 \,.
\ee
It follows that \eq{++ExSp17} will hold if and only if
\bel{12XII1654}
 \left(\partial_\beta h^{\beta}{}_\alpha-\frac 12 \partial_\alpha h^\beta{}_\beta\right)\Big|_{t=0}
  =0 =
 \partial_0\left(\partial_\beta h^{\beta}{}_\alpha-\frac 12 \partial_\alpha h^\beta{}_\beta\right)\Big|_{t=0}
 \,.
\ee
Equivalently, taking \eq{++ExSp23} into account,
\beal{12XII16.7}
&
 \partial_0 (h_{00} + h^i{}_{i})|_{t=0}
  =
 2 \partial_i h^i{}_0|_{t=0}
 \,,
&
\\
&
 \label{12XII16.6}
 \partial_0 h_{0i}|_{t=0}
  =
 \big(   \partial_j h^j{}_i +  \frac 12\partial_i(h_{00}- h^j{}_{j})\big)\big|_{t=0}
 \,,
&
\\
 \label{12XII16.9}
&
 \Delta h^i{}_{i}|_{t=0}
  =
 \partial_i \partial_j h^{ij}|_{t=0}
 \,,
&
\\
 &
 \partial_j\big(\partial_0 h^{j}{}_i- \partial_ 0 h^k{}_k \delta^j_i\big) \big|_{t=0}
  = \big(\Delta h_{0i} - \partial_i \partial_j h^j{}_0\big) \big|_{t=0}
 \,.
 &
\eeal{12XII16.5}
The last two equations are of course the linearisations of the usual scalar and vector constraint equations.

There remains  the freedom of choosing $\zeta_\alpha |_{t=0}$ and $ \partial_t\zeta_\alpha |_{t=0}$. We choose
\begin{eqnarray}
 \nn
   &
    \big(\partial_0 h^k{}_{k} - 2 \partial_k h^k{}_{0} - 2\Delta \zeta_0 \big) \big|_{t=0} =0
   \,,
   &
\\
 \nn
   &
    \big(h_{00} + 2\partial_0 \zeta_0 \big) \big|_{t=0} =0
   \,,
   &
\\
 \nn
   &
    \big(h_{0i} +  \partial_i \zeta_0 +  \partial_0 \zeta_i \big) \big|_{t=0} =0
   \,,
   &
\\
   &
   D_i \big(h^i{}_{j} - \frac 13 h^k{}_k \delta^i_j +  D^i \zeta_j +  D_j \zeta^ i - \frac 23 D^k \zeta_k \delta^i_j \big) \big|_{t=0} =0
   \,,
   &
  \label{14XII16.1}
\end{eqnarray}
%
{where $D_i \equiv D^i \equiv \partial_i$ in Cartesian coordinates.}
Indeed, given any $h_{\mu\nu}$ and  $\partial_0h_{\mu\nu} |_{t=0}$, the first equation can be solved for $\zeta_0 |_{t=0}$ under suitable natural conditions on the data; the second defines $\partial_0 \zeta_0|_{t=0}$;
the third defines $\partial_0 \zeta_i|_{t=0}$; finally, the last equation is an elliptic equation for the vector field $  \zeta_i|_{t=0}$ which can be solved~\cite{christodoulou:murchadha}  if one assumes that the field
\bel{12XII16.8}
   \partial_i \big(h^i{}_{j} - \frac 13 h^k{}_k \delta^i_j  \big) \big|_{t=0}
\ee
belongs to a suitable weighted Sobolev or H\"older space, the precise requirements being irrelevant for our purposes.
We simply note that if some components of $h_{ij}$ behave as $1/r$, then $\zeta$ will behave like $\ln r $ in general, which is likely to introduce $\ln r/r$ terms in the gauge-transformed metric.
After performing this gauge-transformation, we end up with a tensor field $h_{\mu\nu}$ which  satisfies
\bel{14XII16.3}
 \partial_0 h^k {}_{k}\big|_{t=0}  =
 h_{00}\big|_{t=0}  =
  h_{0i}\big|_{t=0}  =  \partial_i \big(h^i{}_{j} - \frac 13 h^k{}_k \delta^i_j  \big)\big|_{t=0}   =0
 \,.
\ee
Inserting this into \eq{12XII16.7}-\eq{12XII16.5} we find
\beal{12XII16.7+}
&
 \partial_0  h_{00} |_{t=0}
  = 0
 \,,
&
\\
&
 \label{12XII16.6+}
 \partial_0 h_{0i}|_{t=0}
  =
 -  \frac 16 \partial_i  h^j{}_{j} |_{t=0}
 \,,
&
\\
 \label{12XII16.9+}
&
 \Delta h^i{}_{i}|_{t=0}
  =
0
 \,,
&
\\
 &
 \partial_j\big(\partial_0 h^{j}{}_i- \partial_ 0 h^k{}_k \delta^j_i\big) \big|_{t=0}
  = 0
 \,.
 &
\eeal{12XII16.5+}
The further requirement that $h^i{}_i$ goes to zero as $r$ tends to infinity together with the maximum principle  gives
\bel{25XII16.1}
 h^i{}_i |_{t=0}
  =
0
 \,.
\ee
We conclude (compare~\cite{ADMGoldie}) that at any given time $t=t_0$ every linearised gravitational initial data set $(h_{\mu\nu},\partial_t h_{\mu\nu})|_{t=t_0}$ can be  gauge-transformed to the TT-gauge:
writing $k_{ij}=\partial_0 h_{ij}$, we have
\bel{14XII16.5}
 h^{k}{}_k |_{t=t_0} = \partial_i h^i{}_j  |_{t=t_0} =
 k^{k}{}_k  = \partial_i k^i{}_j   = 0
 \,.
\ee
From what has been said and from uniqueness of solutions of the wave equation we also see that in this gauge we will have for all $t$
\bel{14XII16.6}
 h_{00} =  h_{0i} = h^{k}{}_k  = \partial_i h^i{}_j    = 0
 \,,
\ee
which further implies that \eq{14XII16.5} is preserved by evolution.

It should be pointed out that when the construction is carried out on the complement of a ball, e.g. because sources are present, or because we perform the construction at large distances only where the non-linearities become negligible, then \eq{25XII16.1} will not hold in general, and the trace of $h_{ij}$ will be non-trivial,
with the usual expansion in terms of inverse powers of $r$, starting with $1/r$-terms associated with the total mass of the configuration. In such cases our construction below still applies to the transverse-traceless part of the metric.

\subsection{Third order potentials}
 \label{ss17XII16.1}

We will need the following result from~\cite{BeigTT}, which can be summarised as follows: Let $h_{ij}$ be a symmetric, transverse and traceless  tensor on $\R^3$,
\bel{17XII16.11}
 \partial_i h^{i}{}_j = 0 = h^i{}_i
 \,.
\ee
Then there exists a symmetric traceless ``third order potential'' $u_{ij}$ such that
\bel{17XII16.12}
 h_{m\ell} = P(u)_{m\ell}
 \,,
\ee
where
(here $g_{ij}$ denotes the Euclidean metric and $D^i$ the associated covariant derivative)
\bel{17XII16.13}
 P(u)_{m\ell}:= \frac 12 \epsilon_m{}^{ij} \partial_i \big( \Delta  u_{j\ell} -
  2  \partial_{(\ell} D^n u_{j) n} + \frac 12 g_{j\ell} D^n D^k u_{n k}
    \big)
 \,,
\ee
and where $u = u_{ij}dx^i dx^j$ can be constructed by the following procedure: Letting
\bel{17XII16.14}
 \sigma _{ijk}(\vec x):=  \int_0^1 \epsilon_{ij}{}^\ell h_{\ell k}(\lambda \vec x) \lambda (1 - \lambda)^2 d\lambda
 \,,
\ee
we set
\bel{17XII16.15}
u_{j\ell} = 2 x^m x^n x_{(j} \sigma_{\ell)mn} + r^2 x^m \sigma_{m(j\ell)}
 \, .
\ee
(This is clearly symmetric, and tracelessness is not very difficult to check. Other third-order-potentials $u$ are possible, differing by an element of the kernel of $P$.)
One way to see how  (\ref{17XII16.13}) arises is to note that $P(u)$ is, apart from a numerical factor, the linearisation at the flat metric of the Cotton-York tensor
in the direction of the trace-free tensor $u$.  For   (\ref{17XII16.15}), the formulae follow   by successively  integrating thrice
the two-forms given in~\cite{BeigTT}, at each step using the Poincar\'e formula \eq{19XII16.1} below. We sketch the construction in Appendix~\ref{sA17XII16.2}.

The converse is also true: given any symmetric trace-free tensor $u_{ij}$, the tensor field $P(u)$ defined by \eq{17XII16.13} is symmetric, transverse and traceless (of which only the last property and the vanishing of the divergence on the first index are obvious).

As an example, consider $h_{ij}$ describing a plane gravitational wave in TT-gauge propagating in direction $\vec k$,
\bel{19XII16.3}
 h_{ij} (\vec k) = \Re \big( H_{ij} e^{i\vec k \cdot \vec x}\big)
 \,,
 \quad
  \partial_\ell H_{ij}=0=H^i{}_i=H_{ij}k^j
 \,,
\ee
with possibly complex coefficients $H_{ij}$, where $\Re$ denotes the real part .
Then
\bean
 \sigma_{ijk} & = & \Re\big(
  \epsilon_{ij\ell} H^\ell{}_k
   \int_0^1 e^{i\lambda \vec k \cdot \vec x} (\lambda - 2\lambda^2 + \lambda^3) d\lambda \big)
\\
 \label{19XII16.2}
 & = &
  \Re\big(W \epsilon_{ij\ell} H^\ell{}_k\big) \,,\quad
  \mbox{where}
\\
 \label{25XII16.2}
    W(\vec x) & = &
  \frac{2 i e^{i \vec k \cdot \vec x}
   (\vec k \cdot \vec x+3
   i)-\vec k \cdot \vec x
   (\vec k \cdot \vec x-4
   i)+6}{(\vec k \cdot \vec x)^4}
\\
 &&
 \nn
   \mbox{(which tends to $1/12$ when $\vec k \cdot \vec x$ tends to zero)}
   \,,
\\
u_{j\ell} & = &  \Re\Big( W \big(2 x^m x^i x_{(j}   \epsilon_{\ell)m k} H^k{}_i  - r^2 x^i   \epsilon_{ik (j} H^k{}_{\ell)} \big)
 \Big)
 \,.
\eeal{19XII16.4}

As another example, consider the family of fields
\bel{21XII16.11}
 u_{ij} = \ln (1+ r^2) \big( D_i \lambda_j + D_j \lambda_i - \frac 23 D^k \lambda_k g_{ij} \big)
 \,.
\ee
Tensors of the form \eq{21XII16.11} \emph{with the $\ln (1+r^2)$ term removed} form the kernel of $P$ for any $\lambda_i$ (cf. Appendix~\ref{sA17XII16.2}), which easily implies that if $\lambda \sim O(r^\sigma)$ for large $r$ then $h_{ij} \sim O(r^{\sigma-4})$, for all $\sigma\in \R$.

We also note that if $h_{ij}$ is  compactly supported to start with, then $u_{ij}$ can also be chosen to be compactly supported; compare Appendix~\ref{sA17XII16.2}.

\subsection{Shielding gravitational Cauchy data}
 \label{ss17XII16.2}

We are ready to prove now a somewhat more general version of our previous claim, that \emph{at any given time $t$, and given any region $\Omega\subset \R^3$, every vacuum initial data set for the gravitational field $(h_{ij},k_{ij})$  can be deformed to a new vacuum initial data set $(\tilde h_{ij},\tilde k_{ij})$ which  coincides  with $(h_{ij},k_{ij})$ on $\Omega$ and vanishes  outside of a slightly larger set}.

Indeed, consider any linearised gravitational field in the gauge \eq{14XII16.6}. Denote by $(h_{ij},k_{ij})$  the associated Cauchy data at $t$, and let $(u_{ij},v_{ij})$ denote the corresponding  potentials discussed in Section~\ref{ss17XII16.1}, thus
\bel{17XII16.1}
 (h_{ij},k_{ij}) = (P(u)_{ij},P(v)_{ij})
  \,,
\ee
where $P$ is the third-order differential operator of \eq{17XII16.13}.
Let $\Omega$ be any open subset of $\R^3$ and let $\widetilde \Omega $ be any open set containing $\overline \Omega$. Let $\chi_\Omega$ be any smooth function which is identically equal to one on $\Omega$ and which vanishes outside of $\widetilde \Omega $. Then the initial data set
\bel{17XII16.2}
 (\tilde h_{ij},\tilde k_{ij}) = (P( \chi_{\Omega} u)_{ij},P(\chi_{\Omega} v)_{ij})
\ee
satisfies the vacuum constraint equations everywhere, coincides with  $(  h_{ij}, k_{ij}) $ in $\Omega$ and vanishes outside of $\widetilde \Omega $.

When $\Omega$ is bounded, the new fields $ (\tilde h_{ij},\tilde k_{ij}) $ can clearly be chosen to vanish outside of a bounded set. For example, consider a plane wave solution as in \eq{19XII16.3}. Multiplying the potentials \eq{19XII16.4} by a cut-off function $\chi_{B(R_1)}$ which equals one on $B(R_1)$ and vanishes outside of $B(R_2)$ provides compactly supported gravitational data which coincide with the plane-wave ones in $B(R_1)$. (Alternatively one can replace $\vec k \cdot \vec x$ in the first line of \eq{19XII16.2}, or in \eq{25XII16.2}-\eq{19XII16.4}, by $\vec k \cdot \vec x \ \chi_{B(R_1)}$.) In the limit $\vec k =0$, so that $h_{ij}$ is constant and, e.g., $k_{ij}=0$, one obtains data which are Minkowskian in $B(R_1)$, and outside of $B(R_2)$, and describe a burst of radiation localised in a spherical shell.  Note that the Minkowskian coordinates for the interior region are distinct from the ones for the outside region. The closest full-theory configuration to this would be  Bartnik's time symmetric initial data set~\cite{Bartnik93} which are flat inside a ball of radius $R_1$, and which can be Corvino-Schoen deformed to be Schwarzschildean outside of the ball of radius $R_2$; here $R_2$ will be much larger than $R_1$ in general, but can be made as close to $R_1$ as desired by making the free data available in Bartnik's construction sufficiently small.

 For $\Omega$'s which are not bounded it is interesting to enquire about fall-off properties of the shielded field. This will depend upon the geometry of $\Omega $ and the fall-off of the initial field:

For cone-like geometries, as considered in~\cite{CarlottoSchoen,ChDelayExoticAERiemannian}, and with $h_{\mu\nu}=O(1/r)$, the gravitational field in the screening region will fall-off again as  $O(1/r)$.
This is rather surprising, as the gluing approach of~\cite{CarlottoSchoen} leads to a loss of decay even for the linear problem. One should, however, keep in mind that the transition to the TT-gauge for a metric which falls-off as $1/r$ is likely to introduce $\ln r/r$ terms in the transformed metric, which will then propagate to the gluing region.

As another example, consider the set $\Omega= (a,b)\times \R^2$, which is not covered by the methods of~\cite{CarlottoSchoen}. Our procedure in this case applies  but if $h_{\mu\nu}=O(1/r)$, and if the cut-off function  is taken to depend only upon the first variable of the product $\Omega= (a,b)\times \R^2$, one obtains a gravitational field $\tilde h_{\mu\nu}$ vanishing outside a slab $\widetilde \Omega = (c,d)\times \R^2$, with $[a,b]\subset (c,d)$,  which might grow as $r^2\ln r$ when receding to infinity within the slab.

So far we have been concentrating on ``shielding''. But of course the above can be used to glue linearised field across a gluing region, by interpolating the respective $u$'s to each other in the gluing zone. Equivalently, screen each of the fields which are glued  to zero across the gluing region, and add the resulting new fields.

\section{Shielding Maxwell fields}
 \label{s12XII16.1}

Maxwell equations in Minkowski space-time share at least two features with linearised gravity: existence of constraint equations, and existence of gauge transformations. It might therefore be unsurprising that there exists a version of the Carlotto-Schoen construction which applies to the Maxwell equations; compare~\cite{ErwannTT,CorvinoPollack} for a discussion of the Maxwell equivalent of the Corvino-Schoen construction, which generalises without further ado to the Carlotto-Schoen setting.  We wish to show here how to carry-out the
\emph{shielding of Maxwell fields with Maxwell fields} in an elementary way.

Recall that solutions of the source-free Maxwell equations are in one-to-one correspondence with their initial data at time $t$; these are simply the electric field $\vec E$ and the magnetic field $\vec B$ at $t$. These fields are not arbitrary, but satisfy the constraints
\bel{17XII16.5}
 \divrm \vec E = \divrm \vec B = 0
 \,.
\ee
On $\R^3$, these imply the existence of vector potentials $\vec \omega$ and $\vec A$ such that
\bel{17XII16.6}
   \vec E = \curl \, \vec\omega\,, \quad  \vec B = \curl \, \vec A
 \,.
\ee
In fact there is an explicit formula for $\vec \omega$,
\bel{17XII16.7}
    \omega_i  = \epsilon_{jik} x^j \int_0^1 E^k (\lambda x) \lambda d\lambda
 \,,
\ee
similarly for $\vec A$. Using \eq{17XII16.6}  it is straightforward to show that \emph{at any given time $t$, and given any region $\Omega\subset \R^3$, every sourceless Maxwell fields $(\vec E, \vec B)$ can be deformed to new sourceless Maxwell fields which coincide with $(\vec E, \vec B)$ on $\Omega$ and vanish   outside a slightly larger set}. Indeed, letting $\widetilde \Omega  $ and $\chi_\Omega$ be as in the paragraph following \eq{17XII16.1}, the new Maxwell fields at $t$
\bel{17XII16.8}
   \vec E = \curl \, (\chi_\Omega \vec\omega) \ \mathrm{and}\ \vec B = \curl \,  (\chi_\Omega \vec A)
\ee
are divergence-free, coincide with the original fields on $\Omega$, and vanish outside of $\widetilde \Omega $.

One can solve the Cauchy problem for the Maxwell equations with the new initial data \eq{17XII16.8} to obtain the associated space-time fields, if desired.

The question then arises%
\footnote{We are grateful to Peter Aichelburg for pointing-out the issue to us.}
if every such configurations can be realised by an experimentalist in the lab. Here ``an experimentalist'' is defined as someone whose laboratory equipment can produce any desired electric charges $\rho$ and currents $\vec j$ subject to the conservation law
\bel{17XII16.9}
 \frac{\partial \rho}{\partial t} +
 \divrm \vec j = 0
 \,.
\ee
These, in turn, will produce Maxwell fields as dictated by the Maxwell equations written in their tensorial special-relativistic form:
\bel{17XII16.10}
 \partial_\nu F^{\mu\nu} = 4 \pi j^\mu
 \,,
 \ \mathrm{where}
 \
 (j^\mu)=(\rho, \vec j)
 \,.
\ee
More specifically,  let us describe the lab as the following ``world-volume'':
$$
 \widetilde \mcU:=[t_0,t_3] \times\widetilde \Omega \subset \R^4
 \,.
$$
The region within the lab where the desired Maxwell fields need to be produced will be the set
$$
 \mcU:=[t_1, t_2] \times\Omega \subset \widetilde{\mcU}
 \,,
$$
with $t_0<t_1\le t_2 <t_3$ and $\overline \Omega \subset \widetilde \Omega$.
Let $F_{\mu\nu}$ be a source-free solution of the Maxwell equations in $\mcU$, as needed to carry out the desired experiments.

The following prescription tells us what are the charges and currents outside of $\mcU$ which will produce a Maxwell field $\tilde F_{\mu\nu}$ coinciding with $F_{\mu\nu}$ in $\mcU$,  out of a vacuum configuration at $t\le t_0$: Let $\chi_\mcU$ be a smooth function which is identically one on $\mcU$ and which vanishes outside of $\widetilde \mcU$. Let $A_\mu$ be any four-vector potential associated with $F_{\mu\nu}$, e.g.
\bel{19XII16.1}
 A_\mu (x^\alpha)= x^\nu \int_0^1 F_{\nu\mu}(\lambda x ^\alpha) \lambda d\lambda
 \,.
\ee
Set $\tilde A_\mu = \chi_{\mcU} A_\mu$,  $\tilde F_{\mu\nu} = \partial_\mu \tilde A_\nu -  \partial_\nu \tilde A_\mu$, and
\bel{17XII16.21}
 j^\mu := \frac 1 {4\pi} \partial_\nu \tilde F^{\mu\nu}
 \,.
\ee
Then $\tilde F_{\alpha\beta}$ vanishes outside of the lab world-volume $\widetilde \mcU$, and coincides with the desired field $F_{\alpha\beta}$ in the world-volume $\mcU$ of the experiment.
If the experimenter can produce the four-current \eq{17XII16.21} with her apparatus, she will be able to create  the desired Maxwell field in the region where the experiment will take place.

It would be of interest to devise an analogous procedure for the gravitational field, keeping in mind the supplementary difficulty of maintaining positivity of energy density.

\section{The Weyl tensor formulation}
 \label{s26II17.2}

As is well-known, the vacuum Einstein equations imply a system of equations for the metric and the Weyl tensor~\cite{GerochEspositoWitten},
\bel{26II17.1}
 \nabla_\mu C^{\mu}{}_{\alpha\beta\gamma} = 0
 \,.
\ee
which implies a symmetrizable-hyperbolic system of equations in dimension $1+3$ (cf., e.g., \cite{F1}).
In the linearised case the equations for the metric and the Weyl tensor decouple, so that one can consider the Weyl tensor equations linearised on Minkowski space-time on their own. We show in Appendix~\ref{s26II17.1} the equivalence of this approach to the metric one, in the sense that a linearised Weyl tensor is always accompanied by a linearised
metric (the reverse property being obvious).

In space-time dimension four \eq{26II17.1} can be rewritten in Maxwell-like form. In this approach (compare~\cite{GerochEspositoWitten}) the evolution equations for two symmetric trace-free tensors $E_{ij}$ and $B_{ij}$,
\bel{26II17.10}
 \partial_t E_{ij} = - \epsilon_{i}{}^{k\ell} \partial_k B_{\ell j}
 \,,
 \quad
 \partial_t B_{ij} = \epsilon_{i}{}^{k\ell} \partial_k E_{\ell j}
 \,,
\ee
are complemented by the constraint equations
\bel{26II17.11}
 D^i E_{ij} = 0 = D^iB_{ij}
 \,.
\ee
Here $E_{ij}$ is the electric part and $B_{ij}$ the magnetic part of the Weyl tensor:
\bel{26II17.12}
 E_{ij} = C_{0i0j}
 \,,
 \quad
 B_{ij} = \star C_{0i0j}
 \,,
\ee
with
$$
\star C_{\alpha\beta\gamma\delta} = \frac 12 \epsilon_{\alpha\beta}{}^{\mu\nu} C_{\mu\nu\gamma\delta}
 \,.
$$
The symmetry and tracelessness of $E_{ij}$,  as well as tracelessness of $B_{ij}$ are obvious from the symmetries of the Weyl tensor.
The symmetry of   $B_{ij}$ follows from the less-obvious double-dual symmetry of the Weyl tensor (cf., e.g., \cite[Proposition~4.1]{ChristodoulouKlainermanCPAM})
$$
  \epsilon_{\alpha\beta}{}^{\mu\nu} C_{\mu\nu\gamma\delta}=
  \epsilon_{\gamma\delta}{}^{\mu\nu} C_{\mu\nu\alpha\beta}
 \,.
$$
We show in Appendix~\ref{s26II17.1} how the vanishing of the divergence of $E_{ij}$ relates to the linearised scalar constraint equation, and how the symmetry of $B_{ij}$ relates to the vector constraint equation.

Since both $ E_{ij}$ and $B_{ij}$ are transverse and traceless, each of them comes with its own third-order potential $u_{ij}$ as described in Section~\ref{ss17XII16.1}, so that shieldings and gluings can be performed on each of them directly, without having to invoke the metric tensor.

\appendix

\section{Integrating 2-forms on $\mathbb{R}^3$}
 \label{sA17XII16.2}

In this Appendix we address the question of asymptotic behaviour of potentials for closed two-forms. The analysis below has obvious generalizations to $p$-forms on $\mathbb{R}^n$ ($n>3$) with $1 < p < n$.

\begin{lemma}
  \label{L21XII16.1}Let $\omega_{ij} (x) = \omega_{[ij]} (x)$ be a closed 2-form on $\mathbb{R}^3$ with $\omega_{ij} = O(r^{\sigma})$, $\alpha \in \R$. Then there exists a 1-form $\omega_i(x)$ with $\partial_{[i}\omega_{j]} = \omega_{ij}$ satisfying $\omega_i(x) = O(r^{1 + \sigma})$ if $\sigma \ne - 2$, $\omega_i(x) = O(r^{- 1} \ln r)$ otherwise.
\end{lemma}

\proof Consider first the case $\sigma \geq - 2$. Then
\be
 \label{21XII16.41}
  \omega_i(x) = 2 \,x^j \int_0^1 \omega_{ji}(\lambda x) \lambda d\lambda = O (r^{1 + \alpha})
  \quad
  \mathrm{when}\,\, \sigma > -2
   \,,
\ee
and $ \omega_i (x) = O (r^{- 1} \ln r)$ when $\sigma = - 2$.
To see this, use spherical coordinates $(r,\theta,\varphi)$ in the argument of $\omega_{ij}$ and substitute $s/r$ for $ \lambda$. When $\sigma < - 2$, consider
\be
   \mu_i(x) = - 2 \,x^j \int_1^\infty \omega_{ji}(\lambda x) \lambda d\lambda
    \,,
\ee
which converges and has the right decay at infinity, but blows up at the origin. The previous expression $\omega_i$ is still defined and, in the shell $\overline{B(2,0)} \setminus B(1,0)$, differs from $\mu_i$ by a closed 1-form. Since this shell is simply connected, this difference $\Delta_i := \omega_i - \mu_i$ satisfies $\Delta_i = \partial_i f$ for some function $f$. Now extend $f$ smoothly
to a function $F$ on all of $B(2,0)$. Then the 1-form given by $\omega_i + \partial_i F$ in the interior and by $\mu_i$ in the exterior satisfies our requirements.
\qed\medskip

An essentially identical argument shows that if $\omega_{ij}$ has compact support, then $\omega_i$ can also be chosen with compact support (which also follows from standard results in algebraic topology~\cite[Corollary~4.7.1]{BottTu}).

\section{Construction of the potential $u$}
 \label{s21XII16.1}

For the convenience of the reader we review the construction in~\cite{BeigTT}, and take this opportunity to correct a minor mistake in the presentation there,
namely the second sentence after (3.12) there.
Let us define
\bel{21XII16.13}
 \tau_{ijk}  := \epsilon_{ij}{}^l h_{lk}
 \,.
\ee
Since $D_{[i}\tau_{jk] l} = \frac 13 \epsilon_{ijk} D _m h^m{}_l= 0$, there exists a tensor field $U_{ij}$ such that
\bel{21XII16.14}
 \tau_{ijk}  := D_{[i} U_{j]k}
 \,.
\ee
Symmetry of $h_{ij}$ implies that all traces of $\tau_{ijk}$ vanish, which implies in turn that
\bel{21XII16.15}
 D_{[l} U_{i}{}^{[k} \delta_{j]}{}^{l]}=0
 \,.
\ee
Hence there exists a tensor field $U_{ijk}$, which can be chosen to be antisymmetric in $jk$, so that
\bel{21XII16.16}
 D_{[i} U_{j]}{}^{kl} + U_{[i}{}^{[k} \delta_{j]}{}^{l]} =0
 \,.
\ee
From tracelessness of $h_{ij}$ one finds $\tau_{[ijk]}=0$, which shows that there exists a vector field $V_i$ such that
\bel{21XII16.17}
 - \frac{1}{3} U_{[jk]} + D_{[j} V_{k]} =0
 \,.
\ee
\Eqs{21XII16.16}{21XII16.17} together with some algebra give
\beqa
    D_{[l}\big(2 U_{ij]}{}^k - 3 V_{i} \delta_{j]}{}^k
\big) =0
\,,
\eeqa
which implies existence of a potential $V_{ij}$:
\beqa
 \label{21XII16.18}
\frac{2}{3} U_{[ij]}{}^k - V_{[i} \delta_{j]}{}^k + D_{[i} V_{j]}{}^k = 0
 \,.
\eeqa
Setting
$$
 u_{ij} := - 3 V_{(ij)} + \delta_{ij} V_k{}^k
 \,,
$$
a lengthy calculation shows that
\beqa
U_{ij} &=& 3D_i V_j + \frac{1}{2} g_{ij} D^k D^l u_{kl} + \Delta u_{ij}
\nn \\
&& \mbox{} - 2 D^k D_{(i} u_{j)k} - D_i D_j V_d{}^d \,.
\eeqa
Thus neither $V_{[ij]}$, nor $V_i{}^i$, nor $V_i$ contribute to $D_{[i} U_{j]k}$ and we finally obtain (\ref{17XII16.13}).
For (\ref{17XII16.14}), we have to successively write down expressions for (i) $U_{ij}$, (ii) $(U_{ijk},V_i)$, and (iii) $V_{ij}$, at each step using formula
(A.1), and take the symmetric, tracefree part of $-3 V_{ij}$ at the end. In going from (i) to (ii) and (ii) to (iii) one uses the identities
\beqa \int_0^1\!\!\!\int_0^1 \!\!F(\lambda \lambda' x) \lambda \lambda'^2 d\lambda d \lambda' = \int_0^1 \!\! F(\lambda x) \lambda(1 - \lambda) d\lambda
 \eeqa
and
\beqa
\int_0^1\!\!\!\int_0^1 \!\!F(\lambda \lambda' x) \lambda (1 - \lambda) \lambda'^3 d\lambda d \lambda' = \int_0^1 \!\! F(\lambda x) \lambda \frac{(1 - \lambda)^2}{2} d\lambda\,,
 \eeqa
respectively. The rest is index gymnastics.

If $h_{ij}= O(r^\sigma)$ for large $r$, from what has been said here and in Section~\ref{sA17XII16.2}, or by analysing \eq{17XII16.14} for $\sigma \ge -4$, we find that $u_{ij}$ can be chosen to be of $O(r^{\sigma+3})$ when
$\sigma \not \in \{- 4, -3, -2\}$, and $u_{ij}$ of $O(r^{\sigma + 3} \ln r)$ otherwise.  Furthermore, if $h_{ij}$ is compactly supported, then $u_{ij}$ can also be chosen to be compactly supported.

We end this Appendix with an analysis of the kernel of $P$ on a simply connected region. For this we follow through the steps starting from \eq{21XII16.14} with  $\tau_{ijk} = 0$, which implies existence of a potential $M_i$ such that
\begin{equation}
U_{ij} = D_i M_j
 \,.
\end{equation}
Next, from \eq{21XII16.16}, there exists an antisymmetric tensor field $M^{kl} = M^{[kl]}$ such that
\begin{equation}
U_j{}^{kl} + M^{[k} \delta_{j}{}^{l]} = D_j M^{kl}
 \,.
\end{equation}
Equation~\eqref{21XII16.17} implies the existence of a function $\phi$ such that
\begin{equation}
V_i - \frac{1}{3} M_i = D_i \phi
 \,.
\end{equation}
Inserting into \eq{21XII16.18} we find that
the terms involving $M_i$ cancel so that
\begin{equation}
D_{[i}\left( \frac{2}{3} M_{j]}{}^k + V_{j]}{}^k - \phi \,\delta_{j]}{}^k)\right) = 0
 \,.
\end{equation}
Consequently
\begin{equation}
V_{ij} = - \frac{2}{3} M_{ij} + \phi\, \delta_{ij} + D_i N_j\,,
\end{equation}
so that
\begin{equation}
V_{(ij)} - \frac{1}{3} V_k{}^k = D_{(i} N_{j)} - \frac{1}{3} D_k N^k\,.
\end{equation}
Setting $\lambda_i = -3 N_i/2$, we conclude that any tensor field satisfying $P(u)=0$ on a simply connected region can be written as
\bel{25XII16.11}
 u_{ij} =  D_i \lambda_j + D_j \lambda_i - \frac 23 D^k \lambda_k g_{ij}
  \,.
\ee

\section{A potential for the linearised Riemann tensor}
 \label{s26II17.1}

In this appendix we show that every linearised Riemann tensor on a star-shaped subset of $\R^d$ arises from a linearised metric $h_{\mu\nu}$, in arbitrary dimension $>2$, where $h_{\mu\nu}$ is defined uniquely up to the usual gauge transformations. We leave it as an exercise to the reader to obtain an explicit formula for $h_{\mu\nu}$ by following the steps of our calculation below.

Suppose, thus, that $R_{\mu\nu\rho\sigma}$ is a field on Minkowski space-time  having the algebraic symmetries of the Riemann tensor and satisfying the Bianchi identity $\partial_{[\mu} R_{\nu\rho]\sigma\tau} = 0$. Then
\begin{equation}
R_{\mu\nu\rho\sigma} = \partial_{[\mu} F_{\nu] \rho \sigma}
 \label{26II17.1}
\end{equation}
with $F_{\mu\nu\rho} = F_{\mu[\nu\rho]}$. But, since $R_{[\mu\nu\rho]\sigma} = 0$,
\begin{equation}
F_{[\mu\nu]\rho} = \partial_{[\mu} H_{\nu]\rho}\,.
 \label{26II17.2}
\end{equation}
Inserting the identity
\begin{equation}
 F_{\nu\rho\sigma} = F_{[\sigma\nu]\rho} + F_{[\sigma\rho]\nu} - F_{[\rho\nu]\sigma}
 \label{26II17.3}
\end{equation}
into \eqref{26II17.2}, and the resulting equation into \eqref{26II17.1}, we find the identity:
\begin{equation}
R_{\mu\nu\rho\sigma} = 2 \,\partial_{[\mu} h_{\nu][\rho, \sigma]}\,,
 \label{26II17.4}
\end{equation}
where
$$
 h_{\mu\nu} = H_{(\mu\nu)}
 \,.
$$
The right-hand side of
\eq{26II17.4} multiplied by $\epsilon$ is, up to $O(\epsilon^2) $ terms, the   Riemann tensor of the metric $\eta_{\mu\nu}+\epsilon h_{\mu\nu}$. Equivalently, $R_{\mu\nu\rho\sigma}$ is the linearised Riemann tensor associated with $h_{\mu\nu}$.

The addition of a pure-trace tensor to $h$ does not change the trace-free part of $R_{\mu\nu\rho\sigma}$.
So, for a tensor $C_{\mu\nu\rho\sigma}$ with Weyl-symmetries satisfying
$\partial_{[\mu} C_{\nu\rho]\sigma\tau} = 0$, there exists a second-order potential $h_{\mu\nu}$ as in \eqref{26II17.4}, which is trace-free.

It is instructive to show equivalence of \eq{26II17.1} to the metric formulation of the theory. For this we note that, in space-time dimension four, \eq{26II17.1} is equivalent to~\cite[Proposition~4.3]{ChristodoulouKlainermanCPAM}
\bel{28II17.1}
 \partial_{[\alpha} C_{\beta\gamma]\mu\nu} = 0
 \,.
\ee
As already pointed out, this implies existence of a symmetric tensor field $h_{\mu\nu}$ such that
\begin{equation}
C_{\mu\nu\rho\sigma} = 2 \,\partial_{[\mu} h_{\nu][\rho, \sigma]}\,.
 \label{26II17.4+}
\end{equation}
But the right-hand side of \eq{26II17.4+} is the linearised Riemann tensor associated with the linearised metric perturbation $h_{\mu\nu}$. Since the left-hand side of  \eq{26II17.4+} has vanishing traces, we conclude that the linearised Ricci tensor associated with $h_{\mu\nu}$ vanishes. Equivalently, $h_{\mu\nu}$ satisfies the linearised Einstein equations.

To understand the nature of the divergence constraint $D^i E_{ij}=0$, let us denote by $r_{ijkl}$ the linearised Riemann tensor of the three-dimensional metric $\delta_{ij}+h_{ij}$, with the associated linearised Ricci tensor $r_{ij}=r^k{}_{ikj}$. We have just seen that $C_{\alpha\beta\gamma\delta}=R_{\alpha\beta\gamma\delta}$ for solutions of $\partial_\alpha C^\alpha{}_{\beta\gamma\delta}=0$, which gives for such solutions
\bel{28II17.5}
 0 = R_{ij} = R^\alpha{}_{i\alpha j} = C^\alpha{}_{i\alpha j} = -C_{0i0j}+r_{ij} = -E_{ij} +r_{ij}
 \,.
\ee
Here we have used the fact that the three-dimensional Riemann tensor differs from the four-dimensional one by quadratic terms in the extrinsic curvature, hence both tensors coincide when linearised at Minkowski space-time. The vanishing of the divergence of the Einstein tensor implies
$$
 D^i r_{ij} = \frac 12 D_j r
 \,,
$$
which together with \eqref{28II17.5} shows that the constraint equation $D^i E_{ij}=0$ is, for asymptotically flat solutions, equivalent to the linearised scalar constraint $r=0$.

Let us show that symmetry of $B_{ij}$ is equivalent to the vector constraint equation. For this let
$$
 k_{ij} = \frac 12 (\partial_0 h_{ij} - \partial_i h_{0j} - \partial_j h_{0i})
$$
denote the linearised extrinsic curvature tensor of the slices $t=\const$. By a direct calculation, or by linearising the relevant embedding equations, we find
\bel{28II17.6}
 R_{0ij\ell} = \partial_\ell k_{ij} - \partial_j k_{i\ell}
 \,.
\ee
Again for solutions of $\partial_\alpha C^\alpha{}_{\beta \gamma \delta} = 0$ it holds that
\bean
 \epsilon^{n\ell m} B_{\ell m} & =  & \frac 12 \epsilon^{n\ell m} \epsilon_{m r s} C_{0\ell}{}^{rs}
  = \frac 12 \epsilon^{n\ell m} \epsilon_{m r s} R_{0\ell}{}^{rs}
  = 2 \delta^{[n  }_{  r }\delta^{ \ell ]}_{  s} D^s k_\ell {}^{r}
\\
 & = &
 D^\ell (k_\ell {}^n - k^m{}_m \delta_\ell^n)
 \,,
\eeal{28II17.7}
as claimed.

Let us finally  consider the kernel of the map sending $h_{\mu\nu}$ into $R_{\mu\nu\rho\sigma}$. Namely, when $R_{\mu\nu\rho\sigma} = 0$,
from \eq{26II17.4} we infer
\begin{equation}
h_{\mu[\nu,\rho]} = \partial_\mu A_{\nu\rho}\,,
 \label{26II17.5}
\end{equation}
where $A_{\nu\rho} = A_{[\nu\rho]}$. But, since $\partial_{[\mu} A_{\nu\rho]} = 0$,
\begin{equation}
A_{\mu\nu} = \partial_{[\mu} B_{\nu]} \,.
 \label{26II17.6}
\end{equation}
Now defining $k_{\mu\nu} = h_{\mu\nu} + \partial_\mu B_\nu$, there results
\begin{equation}
k_{\mu[\nu,\rho]} = h_{\mu[\nu,\rho]} + \partial_\mu \partial_{[\rho} B_{\nu]} = 0\,,
 \label{26II17.7}
\end{equation}
so that $k_{\mu\nu} = \partial_\mu D_\nu$, whence $h_{\mu\nu} = \partial_\mu (D_\nu - B_\nu)$. Finally, using the symmetry of $h_{\mu\nu}$, it follows that
\begin{equation}
h_{\mu\nu} = \partial_{(\mu} \Lambda_{\nu)}
 \label{26II17.8}
\end{equation}
with $\Lambda_\mu = D_\mu - B_\mu$.

\bigskip

\noindent{\sc Acknowledgements:} Supported in part by the Austrian Science Fund (FWF)  project P29517-N16.  Useful discussions with Peter Aichelburg and J\'er\'emie Joudioux are gratefully acknowledged.

\bibliographystyle{amsplain}
\bibliography{%
../references/reffile,%
../references/newbiblio,%
../references/hip_bib,%
../references/newbiblio2,%
../references/bibl,%
../references/howard,%
../references/bartnik,%
../references/myGR,%
../references/newbib,%
../references/Energy,%
../references/netbiblio,%
../references/PDE}

\def\polhk#1{\setbox0=\hbox{#1}{\ooalign{\hidewidth
  \lower1.5ex\hbox{`}\hidewidth\crcr\unhbox0}}} \def\cprime{$'$}
  \def\cprime{$'$}
\providecommand{\bysame}{\leavevmode\hbox to3em{\hrulefill}\thinspace}
\providecommand{\MR}{\relax\ifhmode\unskip\space\fi MR }
\providecommand{\MRhref}[2]{%
  \href{http://www.ams.org/mathscinet-getitem?mr=#1}{#2}
}
\providecommand{\href}[2]{#2}
\begin{thebibliography}{10}

\bibitem{ADMGoldie}
R.L. Arnowitt, S.~Deser, and C.W. Misner, \emph{{The dynamics of general
  relativity}}, Gen. Rel. Grav. \textbf{40} (2008), 1997--2027,
  arXiv:gr-qc/0405109.

\bibitem{Bartnik93}
R.~Bartnik, \emph{Quasi-spherical metrics and prescribed scalar curvature},
  Jour.\ Diff.\ Geom. \textbf{37} (1993), 31--71.

\bibitem{BeigTT}
R.~Beig, \emph{T{T}-tensors and conformally flat structures on
  {$3$}-manifolds}, Mathematics of gravitation, Part I (Warsaw, 1996), Banach
  Center Publ., vol.~41, Polish Acad. Sci., Warsaw, 1997, pp.~109--118.
  \MR{MR1466511 (98k:53040)}

\bibitem{BottTu}
R.~Bott and L.W. Tu, \emph{Differential forms in algebraic topology}, Graduate
  Texts in Mathematics, vol.~82, Springer-Verlag, New York-Berlin, 1982.
  \MR{658304}

\bibitem{CarlottoSchoen}
A.~Carlotto and R.~Schoen, \emph{Localizing solutions of the {Einstein}
  constraint equations}, Invent.\ Math. \textbf{205} (2016), 559--615,
  arXiv:1407.4766 [math.AP].

\bibitem{ChristodoulouKlainermanCPAM}
D.~Christodoulou. and S.~Klainerman, \emph{Asymptotic properties of linear
  field equations in {M}inkowski space}, Commun.\ Pure Appl.\ Math. \textbf{43}
  (1990), 137--199.

\bibitem{christodoulou:murchadha}
D.~Christodoulou and N.{\'O} Murchadha, \emph{The boost problem in general
  relativity}, Commun.\ Math.\ Phys. \textbf{80} (1981), 271--300.

\bibitem{ChBourbaki}
P.T. Chru\'{s}ciel, \emph{{Anti-gravity \`a la Carlotto-Schoen}}, S\'eminaire
  Bourbaki \textbf{1120} (2016), 1--24, arXiv:1611.01808 [math.DG].

\bibitem{ChDelayExoticAERiemannian}
P.T. Chru\'{s}ciel and E.~Delay, \emph{On {Carlotto-Schoen-type}
  scalar-curvature gluings},  (2016), arXiv:1611.00893 [math.DG].

\bibitem{CorvinoPollack}
J.~Corvino and D.~Pollack, \emph{Scalar curvature and the {E}instein constraint
  equations}, Surveys in geometric analysis and relativity, Adv. Lect. Math.
  (ALM), vol.~20, Int. Press, Somerville, MA, 2011, arXiv:1102.5050 [math.DG],
  pp.~145--188. \MR{2906924}

\bibitem{ErwannTT}
E.~Delay, \emph{Smooth compactly supported solutions of some underdetermined
  elliptic {PDE}, with gluing applications}, Commun.\ Partial Diff.\ Eq.
  \textbf{37} (2012), 1689--1716, arXiv:1003.0535 [math.FA]. \MR{2971203}

\bibitem{F1}
H.~Friedrich, \emph{On the regular and the asymptotic characteristic initial
  value problem for {E}instein's vacuum field equations}, Proc.\ Roy.\ Soc.\
  London Ser.\ A \textbf{375} (1981), 169--184. \MR{MR618984 (82k:83002)}

\bibitem{GerochEspositoWitten}
R.~Geroch, \emph{Asymptotic structure of space-time}, Asymptotic structure of
  space-time ({P}roc. {S}ympos., {U}niv. {C}incinnati, {C}incinnati, {O}hio,
  1976), Plenum, New York, 1977, pp.~1--105. \MR{MR0484240 (58 \#4166)}

\end{thebibliography}

\end{document}